\documentclass[final,3p,times]{elsarticle}

\usepackage{amssymb}

\usepackage{lineno}
\usepackage{url}

\journal{Physics Letter B}

\begin{document}
\def\Journal#1#2#3#4{{#1} {\bf #2}, #3 (#4)}

\def\NCA{Nuovo Cimento}
\def\NIM{Nucl. Instr. Meth.}
\def\NIMA{{Nucl. Instr. Meth.} A}
\def\NPB{{Nucl. Phys.} B}
\def\NPA{{Nucl. Phys.} A}
\def\PLB{{Phys. Lett.}  B}
\def\PRL{Phys. Rev. Lett.}
\def\PRC{{Phys. Rev.} C}
\def\PRD{{Phys. Rev.} D}
\def\ZPC{{Z. Phys.} C}
\def\JPG{{J. Phys.} G}
\def\CPC{Comput. Phys. Commun.}
\def\EPJ{{Eur. Phys. J.} C}
\def\PR{Phys. Rept.}
\def\PRV{Phys. Rev.}
\def\JHEP{ J. High Energy Phys.}

\newcommand{ \be }{\begin{equation}}
\newcommand{ \ee }{\end{equation}}
\newcommand{ \bea }{\begin{eqnarray}}
\newcommand{ \eea }{\end{eqnarray}}
\newcommand{ \bfig }{\begin{figure}[htpb]}
\newcommand{ \efig }{\end{figure}}
\newcommand{ \bmn }{\begin{minipage}}
\newcommand{ \emn }{\end{minipage}}
\newcommand{ \bt }{\begin{table}[htpb]}
\newcommand{ \et }{\end{table}}
\newcommand{ \la }{\langle}
\newcommand{ \ra }{\rangle}
\newcommand{ \eps }{{\varepsilon}}
\newcommand{ \rar }{\rightarrow}
\newcommand{ \lar }{\leftarrow}
\newcommand{ \as }{$\alpha_{s}$~}
\newcommand{ \dpt }{$\Delta p_{T}$~}
\newcommand{ \pt }{$p_{T}$~}
\newcommand{ \pT }{$p_{T}$~}
\newcommand{ \mt }{$m_{T}$~}
\newcommand{ \nch }{$n_{ch}$~}
\newcommand{ \dzero }{$D^{0}$}
\newcommand{ \vv }{$v_{2}$~}
\newcommand{\sqrts}{\mbox{$\sqrt{s}$}~}
\newcommand{\sNN}{\mbox{$\sqrt{s_{_{\mathrm{NN}}}}$}~}
\newcommand{\dAu}{\textit{d}+Au}
\newcommand{\CuCu}{Cu+Cu}
\newcommand{\pAu}{\textit{p}+Au}
\newcommand{\AuAu}{Au+Au}
\newcommand{\PbPb}{Pb+Pb}
\newcommand{\pA}{\mbox{\textit{p}+A}}
\renewcommand{\AA}{\mbox{A+A}}
\newcommand{\ppbar}{\mbox{$p\bar{p}$}}
\newcommand{\pp}{\mbox{\textit{p}+\textit{p}}~}
\newcommand{\eeh}{\mbox{$(e^{+}$+$e^{-})/2$}}
\newcommand{\meanpt}{\mbox{$\langle p_T \rangle$}}
\newcommand{\Et}{\mbox{$E_T$}}
\newcommand{\mev}{\mbox{$\mathrm{MeV}$}~}
\newcommand{\gev}{\mbox{$\mathrm{GeV}$}~}
\newcommand{\gevcc}{\mbox{$\mathrm{GeV/}c^2$}~}
\newcommand{\mevcc}{\mbox{$\mathrm{MeV/}c^2$}~}
\newcommand{\gevc}{\mbox{${\mathrm{GeV/}}c$}~}
\newcommand{\mevc}{\mbox{${\mathrm{MeV/}}c$}~}
\newcommand{\TAA}{\mbox{$\mathrm{T}_{AA}$}~}
\newcommand{\raa}{\mbox{$R_{AA}$}~}
\newcommand{\RAA}{\mbox{$R_{AuAu/dAu}$}~}
\newcommand{\RCP}{\mbox{$R_{CP}$}~}
\newcommand{\ie}{\mbox{\textit{i.e.}}}
\newcommand{\eg}{\mbox{\textit{e.g.}}}
\newcommand{\dedx}{\mbox{$dE/dx$}~}
\newcommand{\npart}{\mbox{$N_{\mathrm{part}}$}}
\newcommand{\nbin}{\mbox{$N_{\mathrm{bin}}$}}
\newcommand{ \Jpsi }{$J/\psi$~}
\newcommand{ \Upsi }{$\Upsilon$~}
\newcommand{ \pion }{$\pi$~}
\newcommand{ \chpi }{$\pi^{\pm}$~}
\newcommand{ \chK }{$K^{\pm}$~}
\newcommand{\ks}{$K^{0}_{S}$~}
\newcommand{\Ks}{$K^{0}_{S}$~}
\newcommand{\La}{$\Lambda$~}
\newcommand{\aLa}{$\overline{\Lambda}$~}
\newcommand{\Rho}{$\rho$~}
\newcommand{\C}{$c$~}
\newcommand{\T}{$\times$~}
\newcommand{\nsigpi}{n$\sigma_{\pi}$~}
\newcommand{\nsige}{n$\sigma_{e}$~}
\newcommand{\dphi}{$\Delta\phi$~}
\newcommand{\eep}{$e^{+}e^{-}$~}

\begin{frontmatter}



\title{$J/\psi$ production cross section and its dependence on charged-particle multiplicity in \pp collisions at \sqrts= 200 GeV}


\author{
J.~Adam$^{9}$,
L.~Adamczyk$^{1}$,
J.~R.~Adams$^{31}$,
J.~K.~Adkins$^{21}$,
G.~Agakishiev$^{19}$,
M.~M.~Aggarwal$^{33}$,
Z.~Ahammed$^{56}$,
N.~N.~Ajitanand$^{44}$,
I.~Alekseev$^{17,28}$,
D.~M.~Anderson$^{46}$
% ...
}

\author{(STAR Collaboration)}

\address{$^{1}$AGH University of Science and Technology, FPACS, Cracow 30-059, Poland}
% ...
\address{$^{9}$Creighton University, Omaha, Nebraska 68178}
% ...
\address{$^{17}$Alikhanov Institute for Theoretical and Experimental Physics, Moscow 117218, Russia}
% ...
\address{$^{19}$Joint Institute for Nuclear Research, Dubna, 141 980, Russia}
% ...
\address{$^{21}$University of Kentucky, Lexington, Kentucky 40506-0055}
% ...
\address{$^{28}$National Research Nuclear University MEPhI, Moscow 115409, Russia}
% ...
\address{$^{31}$Ohio State University, Columbus, Ohio 43210}
% ...
\address{$^{33}$Panjab University, Chandigarh 160014, India}
% ...
\address{$^{44}$State University of New York, Stony Brook, New York 11794}
% ...
\address{$^{46}$Texas A\&M University, College Station, Texas 77843}
% ...
\address{$^{56}$Variable Energy Cyclotron Centre, Kolkata 700064, India}
% ...

\date{\today}



\begin{abstract}
We present a measurement of inclusive \Jpsi production at mid-rapidity ($|y|<1$) in \pp collisions at a center-of-mass energy of \sqrts= 200 GeV with the STAR experiment at the Relativistic Heavy Ion Collider (RHIC). 
The differential production cross section for \Jpsi as a function of transverse momentum ($p_T$) for $0<p_T<14$ \gevc and the total cross section are reported and compared to calculations from the color evaporation model and the non-relativistic Quantum Chromodynamics model.
The dependence of \Jpsi relative yields in three \pT intervals on charged-particle multiplicity at mid-rapidity is measured for the first time in \pp collisions at \sqrts= 200 GeV and compared with that measured at \sqrts= 7 TeV, PYTHIA8 and EPOS3 Monte Carlo generators, and the Percolation model prediction.
\end{abstract}

\begin{keyword}
quarkonium \sep \pp collisions \sep multiple parton interactions \sep charged-particle multiplicity


\end{keyword}

\end{frontmatter}


\section{Introduction}
\label{introduction}

Quarkonia are bound states of heavy quark-antiquark pairs ($Q\bar{Q}$). 
Their production in \pp collisions can be factorized into hard and soft processes associated with short and long distance strong interactions, respectively \cite{Factorization}. 
The former is related to production of $Q\bar{Q}$ from hard parton scatterings  and can be calculated by perturbative Quantum Chromodynamics (pQCD).
The latter involves evolution of $Q\bar{Q}$ into bound quarkonium states and is usually parameterized by phenomenological models such as the Color-Evaporation Model (CEM), Color Singlet Model, and Non-Relativistic Quantum Chromodynamics (NRQCD) including both color singlet and octet intermediate states (for a recent review see \cite{review}).
With sizable theoretical uncertainties, these model calculations can describe the measurement results of quarkonium production cross sections from the Tevatron and Large Hadron Collider (LHC) experiments \cite{CEM, COCS1,COCS2,COCS3}.
Precise measurements of quarkonium production over a wide kinematic range at RHIC energies can provide new constraints on model calculations and insights into the quarkonium production mechanism.

Recent studies at the LHC have revealed a faster-than-linear increase in \Jpsi and $D$-meson relative yields with charged-particle multiplicity ($n_{ch}$) at mid-rapidity in \pp collisions at \sqrts= 7 TeV~\cite{ALICEjpsievt, ALICED0evt}, suggesting a strong correlation between hard parton scatterings producing heavy flavor particles and soft underlying processes producing all other particles.
By including Multiple-Partonic Interactions (MPI) \cite{MPI1,MPI2,MPI3}, i.e. several interactions at the parton level occurring in a single \pp collision, PYTHIA8~\cite{pythia8} and EPOS3~\cite{epos3} Monte Carlo (MC) generators can produce an increase in relative yields of heavy flavor particles with $n_{ch}$, but underestimate the observed yields at large $n_{ch}$ \cite{ALICED0evt}. 
Additional effects have been suggested to explain the measurement results at the LHC. 
For example, collective expansion implemented in EPOS3 MC generator \cite{epos3hydro} is found to modify the \pT distribution of final state particles in high multiplicity \pp collisions at \sqrts= 7 TeV, producing a faster-than-linear increase for D-meson relative yields at intermediate and high \pT \cite{ALICED0evt}. Such an effect is however expected to be small at RHIC energies. 
On the other hand, the percolation model \cite{perco} produces particles through interactions of color strings, which are more suppressed for soft processes than hard processes due to the different size of the color strings. It can also produce a faster-than-linear increase in relative yields of heavy flavor particles with \nch that is qualitatively consistent with the LHC results. Such an increase is predicted to be similar at different energies by the percolation model.
Measurements of relative yields of heavy flavor particles versus \nch at RHIC energies can help constrain model calculations and provide knowledge of the energy dependence of MPI in \pp collisions.

In this letter, we present new results of inclusive \Jpsi production at mid-rapidity ($|y|<1$) in \pp collisions at \sqrts= 200 GeV with the STAR experiment at RHIC \cite{STAR}. Both the differential production cross section for \Jpsi as a function of transverse momentum ($p_T$) and the total cross section are obtained with higher precision than the previously published results \cite{STARjpsi,PHXjpsi}. The dependence of \Jpsi relative yields on $n_{ch}$ at mid-rapidity is also determined, for the first time, for \pp collisions at RHIC energies. These results are compared to calculations from various theoretical models and MC generators.

\section{STAR experiment and data analysis}
\label{experiment}
The data used in this measurement were collected with minimum-bias (MB) and high-tower (HT) triggers in \pp collisions at \sqrts= 200 GeV by the STAR experiment in 2012. 
The MB triggers select non-single diffractive \pp collisions with a coincidence signal from the two Vertex Position Detectors (VPD)~\cite{vpd} or the two Beam Beam Counters (BBC)~\cite{bbc}.
The VPD and BBC are located on both sides of the \pp collision region and covering the pseudo-rapidity region of $4.4< |\eta|<4.9$ or $3.3<|\eta|<5.0$, respectively. 
The VPD-triggered MB events are analyzed to study the \Jpsi production with $p_T<1.5$ GeV/$c$, since the size of the sample ($\sim$300 million, corresponding to an integrated luminosity of about ~10 nb$^{-1}$) is significantly larger than that of the BBC-triggered MB events (about 2.66 million).
The latter are used to obtain the \nch distribution in MB \pp collisions, since the BBC has a much higher trigger efficiency than the VPD for low multiplicity \pp collisions.
The HT triggers select \pp collisions producing at least one high-\pT particle with large energy deposition in the Barrel Electromagnetic Calorimeter (BEMC)~\cite{bemc}.
Data collected by the HT0 (HT2) trigger with an energy threshold of $E_{T} > 2.6$ (4.2) GeV, corresponding to an integrated luminosity of 1.36 (23.5) $pb^{-1}$, are analyzed to study \Jpsi production with $p_T>1.5$ (4.0) GeV/$c$. 
The vertex positions of \pp collisions along the beam line direction can be reconstructed from TPC tracks ($V_z^{TPC}$) or from VPD signals ($V_z^{VPD}$). A cut of $|V_{z}^{TPC}|<50$ cm is applied to ensure good TPC  acceptance for all the events.  An additional cut of $|V_z^{TPC} - V_z^{VPD}|<6$ cm is applied to reduce the pile-up background from out-of-time collisions for the VPD-triggered MB events.

The main detectors used in the data analysis are the Time Projection Chamber (TPC)~\cite{tpc}, Time-Of-Flight detector (TOF)~\cite{tof}, and BEMC, all with full azimuthal coverage within $|\eta|<1$.
The TPC records trajectories of charged particles with $p_{T}>0.2$ \gevc in a 0.5 T solenoid magnetic field and determines their momenta and ionization energy losses ($dE/dx$). 
Tracks are required to have a maximum distance of the closest approach to the collision point of 1 cm, a minimum of 20 TPC hits (out of a maximum of 45), and a minimum of 11 TPC hits for $dE/dx$ calculation.
The TOF (VPD) provides the stop (start) time of flight information for charged particles from the collision vertices to the TOF, while the BEMC measures electromagnetic energy deposition. 

\Jpsi candidates are reconstructed through the $J/\psi\rightarrow e^{+}e^{-}$ channel, where electrons are identified using the specific energy loss in TPC, $dE/dx$, the velocity ($\beta$) calculated from the path length and time of flight between the collision vertex and TOF, and the ratio between the momentum and energy deposition in the BEMC ($pc/E$) \cite{STARjpsi}. The normalized $dE/dx$ is defined as 
\begin{equation}
n\sigma_{e} = \frac{ln(dE/dx) - ln(dE/dx|_{Bichsel})}{\sigma_{ln(dE/dx)}},
\end{equation}
where $dE/dx$ ($dE/dx|_{Bichsel}$) is the measured (expected) value, and $\sigma_{ln(dE/dx)}$ is one standard deviation of the $ln(dE/dx)$ distribution. The $n\sigma_{e}$ value is 
required to be within (-1.9, 3) for all electron candidates. $|1/\beta - 1|<0.03$ is required for TOF associated electron tracks, and $0.3<pc/E<1.5$ for BEMC-associated candidates with $p_{T}>1$~GeV/$c$.
Both daughters of \Jpsi candidates are required to pass the $n\sigma_{e}$ requirement, and either the $\beta$ or $pc/E$ requirement. 
For HT-triggered events, at least one daughter of \Jpsi candidates must pass the $pc/E$ requirement and have an energy deposition in the BEMC that is higher than the corresponding HT trigger threshold.

\section{\Jpsi production cross section}
\label{jpsixsec}

\begin{figure*}
\begin{center}$  $
\includegraphics[width=1.0\textwidth]{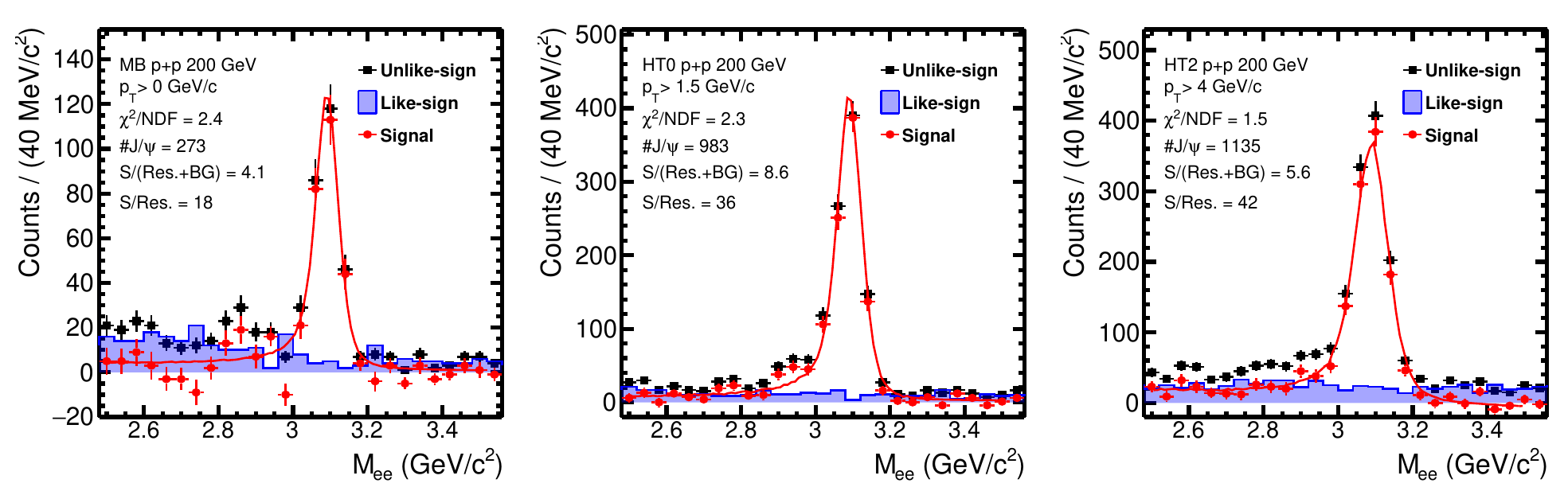}
\caption{Invariant mass distributions for unlike-sign pairs (black squares), like-sign pairs (blue histogram), and difference between unlike- and like-sign  pairs (red circles) from the VPD- (left), HT0- (middle) and HT2-triggered (right) events in \pp collisions at \sqrts= 200 GeV in 2012. The red curve is the combined fit with a Crystal Ball (CB) function and a 1$^\mathrm{st}$-order polynomial function representing signal and residual background, respectively.}\label{fig:jpsisignal}
\end{center}
\end{figure*}

The invariant mass spectra of the reconstructed \Jpsi candidates from different triggered samples are shown in Fig.~\ref{fig:jpsisignal}. The \Jpsi raw yields are extracted by subtracting the invariant mass spectra of like-sign electron pairs ($M_{e^\pm e^\pm}$) from the unlike-sign ones ($M_{e^\pm e^\mp}$). The remaining distribution is fit by a two-component function, composed of a \Jpsi signal distribution, the shape of which is obtained from STAR detector simulation \cite{embed} with the Crystal-Ball function \cite{Crystal}, together with a residual background distribution parameterized by a 1$^\mathrm{st}$-order polynomial function. 
The \Jpsi signal to the residual background ratios in the mass range of $2.9<M_{ee}<3.2$ \gevcc are 18, 36, and 42 for the VPD MB, HT0 and HT2 data, respectively, and have negligible dependence on \nch.
The large values of this ratio reflect that the residual background is small and can be neglected for the measurement of the $n_{ch}$-dependence of $J/\psi$ relative yields.

The \Jpsi production cross sections are obtained by correcting the \Jpsi raw yields for the detector geometric acceptance and efficiency. The vertex finding, track reconstruction, BEMC electron identification, VPD, BBC and HT trigger efficiencies are estimated from detector simulation, while the electron identification efficiencies by TPC $dE/dx$ and TOF $1/\beta$ requirements are estimated from data. 
The cross sections from the VPD MB, HT0 and HT2 data are consistent with each other in the overlapping \pT regions, and are used for $0<p_T<1.5$, $1.5<p_T<4.0$ and $4.0<p_T<14$ GeV/$c$, respectively.
Here unpolarized \Jpsi is used in simulation when calculating the \Jpsi acceptance and efficiency, which is around 20\% for the VPD MB data, 0.6-12\% for the HT0 data, and 1.5-30\% for the HT2 data, respectively. The latter two increase as a function of \Jpsi \pT due to the increasing HT trigger efficiency. 

The total systematic uncertainty for the measured cross section is obtained from the square root of the quadratic sum of the individual systematic uncertainties listed in Table~\ref{tab:yieldSys}, which generally depend on the \Jpsi signal significance in each \pT bin.
The uncertainty in the raw \Jpsi yield extraction, estimated by varying the fitted mass range, by changing the residual background shape from a 1$^\mathrm{st}$-order polynomial function to an exponential function, and by comparing the fit result to the bin counts in $2.9<M_{ee}<3.2$ GeV/$c^2$ corrected for the residual background and for the invariant mass cut efficiency, is between 1-14\%. The uncertainties in the track reconstruction and electron identification efficiencies, estimated by varying the corresponding cuts in data and simulation, are 3-15\% and 4-14\%, respectively. Since the TOF efficiency is calculated from tracks matched with BEMC hits, there is a correction applied to the TOF efficiency by the ratio of true TOF efficiency to that calculated with BEMC-matched tracks. This correction causes an uncertainty of 1-7\%. The HT trigger efficiency uncertainty, estimated by varying the trigger requirement in data and simulation, is 4-13\%. An 8\% normalization uncertainty due to the luminosity determination is applied for both the MB and HT results. An additional 6\% (3\%) normalization uncertainty for the VPD MB (HT) result is added for the VPD (BBC) trigger and vertex reconstruction efficiency, making the total normalization uncertainty 10\% (8.5\%).

\begin{table} \caption{Systematic uncertainties for the measurement of the \Jpsi production cross section. See text for details.}\label{tab:yieldSys}
\centering
\begin{tabular}{cc}

\hline
Type & Uncertainty (\%) \\
\hline
Raw yield extraction & 1-14 \\
Tracking efficiency  & 3-15 \\
Electron identification efficiency & 4-14 \\
TOF-EMC efficiency correlation & 1-7 \\
HT trigger efficiency & 4-13 (HT) \\
Normalization  & 10 (MB) \\ 
               & 8.5 (HT) \\

\hline
\end{tabular}
\end{table}

\begin{figure}
\begin{center}
\includegraphics[width=1.0\textwidth]{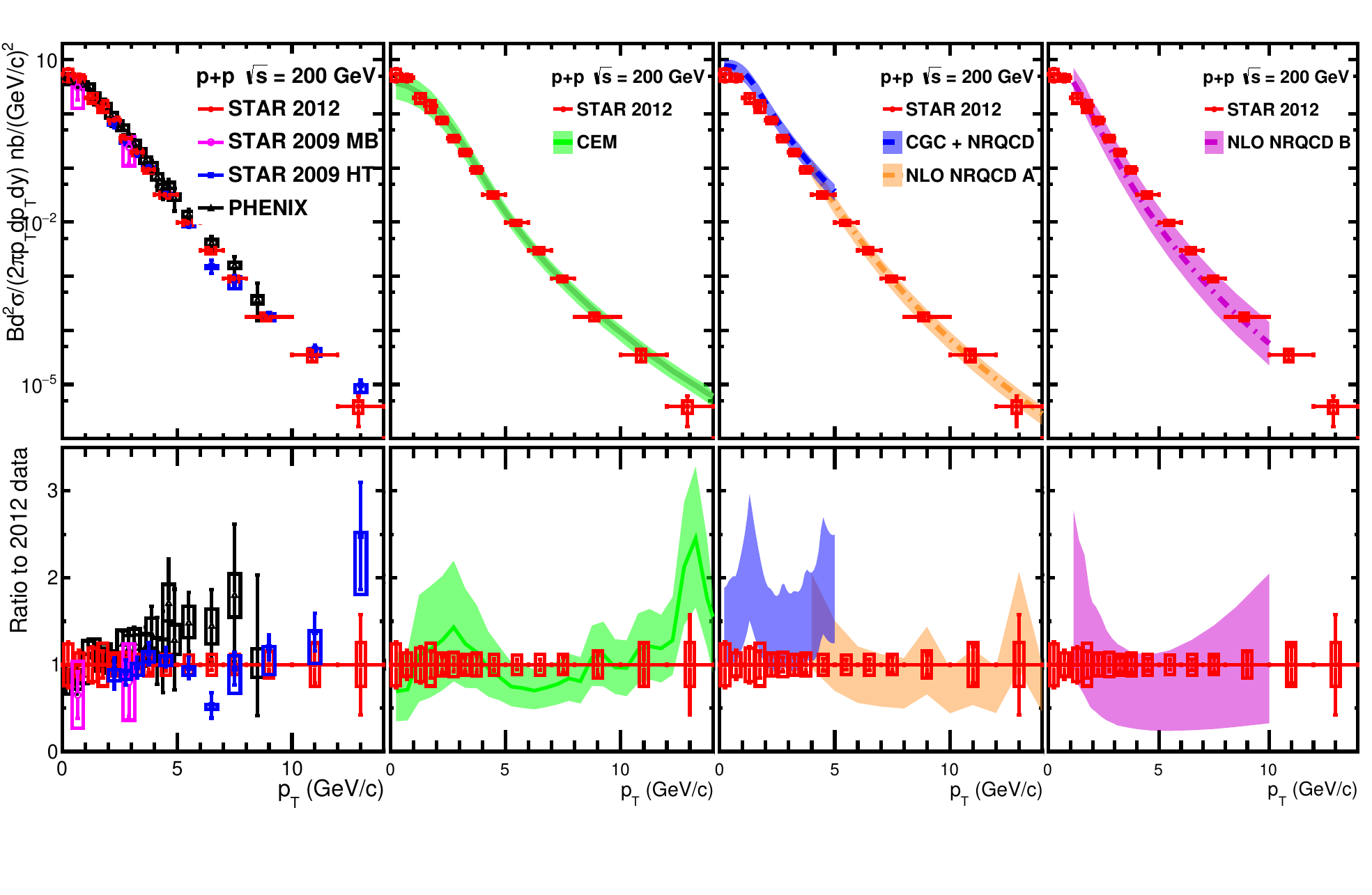}
\caption{(Color online) Top: \Jpsi cross section times branching ratio as a function of \pt in \pp collisions at \sqrts= 200 GeV. Solid circles are from this analysis for $|y|<1$; open circles and blue squares are the published results for $|y|<1$ from STAR~\cite{STARjpsi}; triangles are the published results for $|y|<0.35$ from PHENIX~\cite{PHXjpsi}. Bars and boxes are statistical and systematic uncertainties, respectively. The curves are CEM (green) \cite{CEM}, NLO NRQCD A (orange) \cite{COCS1}, CGC + NRQCD (blue) \cite{CGC2}, and NLO NRQCD B (magenta) \cite{COCS2} theoretical calculations, respectively. Bottom: ratios of these results with respect to the central value from this analysis.}\label{fig:jpsixsec} 
\end{center}
\end{figure}

Figure~\ref{fig:jpsixsec} shows the measured \Jpsi production cross section times the $J/\psi\rightarrow e^+e^-$ branching ratio ($B_{ee}$) as a function of \Jpsi $p_T$. 
The new result is consistent with the published STAR~\cite{STARjpsi} result, but has better statistical precision for $p_T<10$ GeV/$c$. 
It is also consistent with the published PHENIX~\cite{PHXjpsi} result, but has better precision for $p_T>2$ GeV/$c$.
The total \Jpsi production cross section times the branching ratio per rapidity unit is estimated to be
\begin{equation}
B_{ee}\frac{d\sigma}{dy	}|_{y=0} = 43.2 \pm 3.0 (stat.) \pm 7.5 (syst.)~nb.
\end{equation}

Also shown in Fig.~\ref{fig:jpsixsec} are theoretical model calculations. 
The green band represents the result from CEM calculations for $0<p_T<14$ \gevc and $|y|<0.35$ \cite{CEM}, the orange band shows that from Next-to-Leading Order (NLO) NRQCD calculations for $4<p_T<14$ \gevc and $|y|<1$ \cite{COCS1}, 
the blue band depicts the result from NRQCD calculations for $0<p_T<5$ \gevc and $|y|<1$ which incorporates a Color-Glass Condensate (CGC) effective theory framework for small-$x$ resummation \cite{CGC2},
and the magenta band shows that from NLO NRQCD calculations for $1.1<p_T<10$ \gevc and $|y|<0.35$ \cite{COCS2}.
The CEM and NLO NRQCD calculations describe the data reasonably well for the applicable \pT ranges. 
The CGC+NRQCD calculations are consistent with the data within uncertainties, however, the data are close to the lower uncertainty boundary of the theoretical calculation. 
We note that the feed-down contributions from higher charmonium states $\chi_{cJ}$ and $\psi(2S)$ are explicitly considered by the calculations in Ref.~\cite{COCS1, CGC2}, but not by the calculations in Ref.~\cite{CEM, COCS2} where model parameters have been fit to inclusive \Jpsi cross sections from experimental measurements.
We also note that the feed-down contribution from bottom hadron decays is included in the experimental data but not included in any of these calculations, which is predicted to be approximately 10-25\% in the range of $4<p_{T}<14$ \gevc \cite{bfd1,bfd2}. 
As can be seen, except for the two bins at the highest $p_T$, the uncertainties in the experimental results are smaller than those in the theoretical calculations. Therefore, the new STAR result can be used to constrain theoretical model calculations.

\section{Dependence of \Jpsi production on \nch}
\label{JpsiEvtAct}
The BBC-triggered MB data are used to characterize the $n_{ch}$ distribution at mid-pseudorapidity (${|\eta|<1}$) in MB \pp collisions. The \nch distribution for \pp collisions producing \Jpsi is obtained by subtracting the \nch distribution of events containing like-sign electron pairs with $2.9<M_{ee}<3.2$ \gevcc from that of unlike-sign pairs, both of which are reweighted by the inverse of the \Jpsi reconstruction efficiency. Here the raw \nch for each event is obtained from the number of tracks reconstructed in the TPC with a matched hit in the TOF. Unlike the TPC, which is susceptible to pile-up tracks from out-of-time collisions, the TOF only records signals from particles produced from the triggered collision. About 1\% of tracks from out-of-time collisions may randomly match a TOF hit and the effect of these pile-up tracks is considered as a systematic uncertainty. The raw \nch distributions are corrected for the trigger and vertex finding efficiencies, which are important for low multiplicity \pp collisions, and for the TPC and TOF acceptances and efficiencies through an iterative Bayesian unfolding method \cite{unfold}. The response matrix used for unfolding is obtained from MC samples generated with PYTHIA8.183~\cite{pythia8} and convoluted with the detector acceptances and efficiencies. Here the true $n_{ch}$ in the response matrix is defined as the number of charged particles produced promptly from the primary vertex, including pion, kaon, proton, electron, and muon with $|\eta|<1$, $p_{T} > 0$ GeV/$c$, and not from $K^0$ or $\Lambda$ decays. In addition, \Jpsi events require that the $J/\psi$-decay electrons are within $|\eta|<1$ and included in the $n_{ch}$ calculations.

\begin{figure}
\centering
\includegraphics[width=0.75\textwidth]{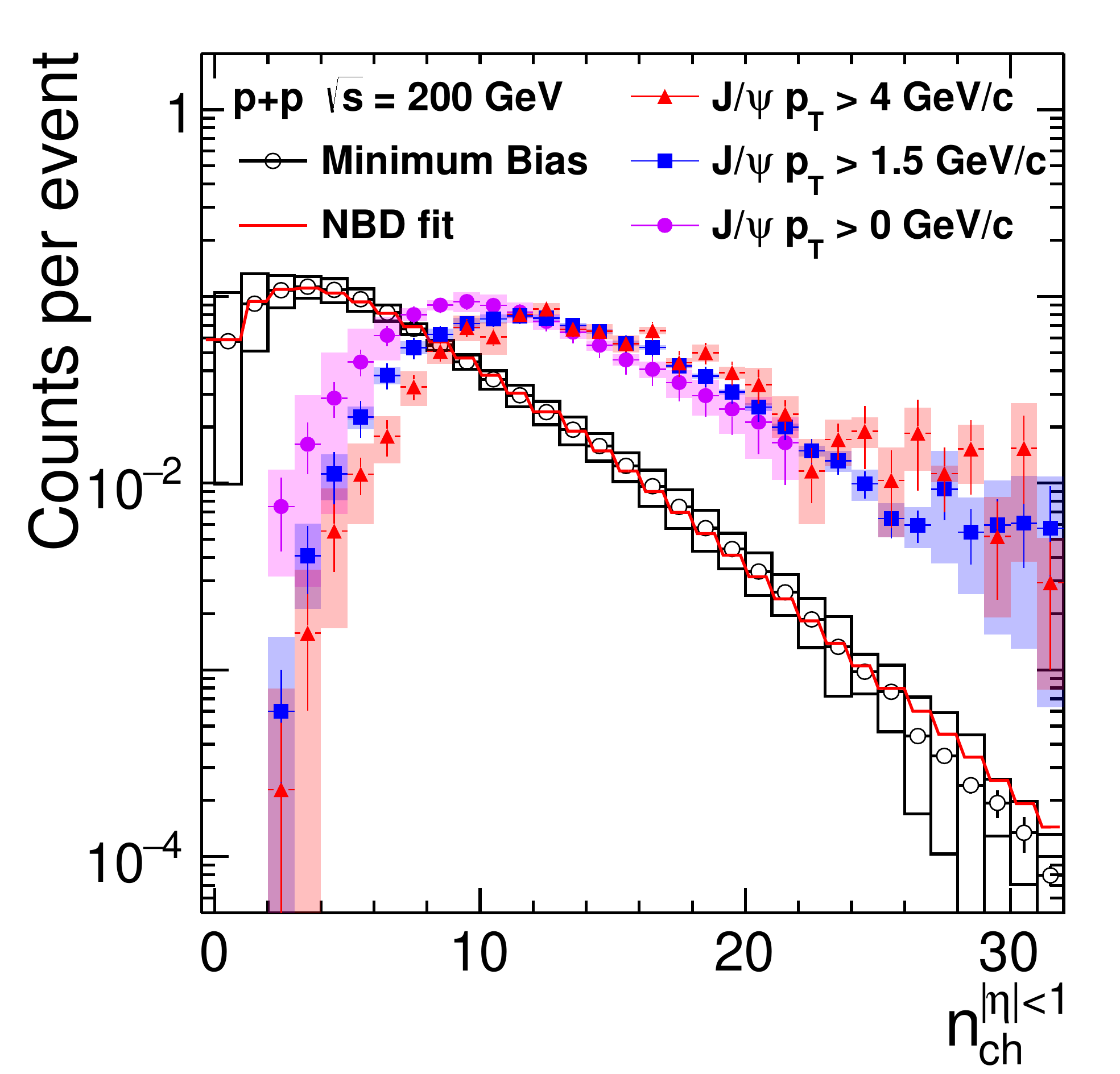}
\caption{(Color online) The corrected $n_{ch}$ distributions at mid-rapidity ($|\eta|<1$) for MB events (open circles) and \Jpsi events with \Jpsi \pt greater than 0 (purple circles), 1.5 (blue squares), and 4 \gevc (red triangles) in \pp collisions at $\sqrt{s}$ = 200 GeV. The fit function is a negative binomial function. Bars and boxes are statistical and systematic uncertainties, respectively. }\label{fig:jpsimult} 
\end{figure}

Figure~\ref{fig:jpsimult} shows the corrected $n_{ch}$ distributions, together with a negative binomial distribution (NBD) function fitted to the distribution in MB \pp collisions. The average charged-particle multiplicity per pseudo-rapidity unit obtained from the fit,  $\langle dN_{ch}^{MB}/d\eta\rangle=2.9 \pm 0.3 (stat.) \pm 0.2 (model) \pm 0.4 (syst.)$, is consistent with the previous STAR published result \cite{embed}. The model uncertainty is estimated from the difference between the average value of the \nch distribution and the NBD fit result. As can be seen, the average $n_{ch}$ for \pp collisions producing $J/\psi$ increases with increasing $J/\psi$ \pT and is higher than that for MB collisions. 

The corrected \nch distributions are divided into five intervals: 0-5, 6-10, 11-15, 16-21 and 22-31. The \Jpsi relative yield $N_{J/\psi}/\langle N_{J/\psi}\rangle$ and relative charged-particle multiplicity $(dN_{ch}^{MB}/d\eta)/\langle dN_{ch}^{MB}/d\eta\rangle$ are estimated for each interval, where $N_{J/\psi}$ is the number of \Jpsi produced per MB collision in a multiplicity interval, and $\langle N_{J/\psi}\rangle$ the average value over the whole multiplicity range. The last bin in MB events is excluded due to the large statistical uncertainty. The systematic uncertainties for $(dN_{ch}^{MB}/d\eta)/\langle dN_{ch}^{MB}/d\eta\rangle$ and $N_{J/\psi}/\langle N_{J/\psi}\rangle$ are summarized in Table~\ref{tab:JpsiEvtActSys}. The vertex finding efficiency correction has 3\% uncertainty, estimated by the difference between the default PYTHIA8 tune and the STAR heavy flavor tune~\cite{hftune}. The track reconstruction efficiency correction is obtained from detector simulation. It depends on \nch and has 4-5 \% uncertainty for MB events and 4-23 \% uncertainty for \Jpsi events. The unfolding process uncertainty is studied by changing the number of Bayesian unfolding iterations and the PYTHIA tune, and by reweighting PYTHIA $n_{ch}$ distribution to match with the unfolded \nch distribution from data. This uncertainty is found to be 2-7 \% and 1-30 \% for MB and \Jpsi events, respectively. The effect of pile-up track contributions is estimated by the difference between the NBD fit and the actual $n_{ch}$ distribution in MB \pp collisions, as well as by the difference in the $n_{ch}$ distribution between the lowest and highest Zero-Degree Calorimeter (ZDC)~\cite{zdc} coincidence rate. The ZDC rate ranged between 1-13 kHz and is proportional to the instantaneous luminosity. The uncertainty due to pile-up track contributions is 2\% and 1-10\% for MB and \Jpsi events, respectively.

\begin{figure*}
\centering
\includegraphics[width=1.0\textwidth]{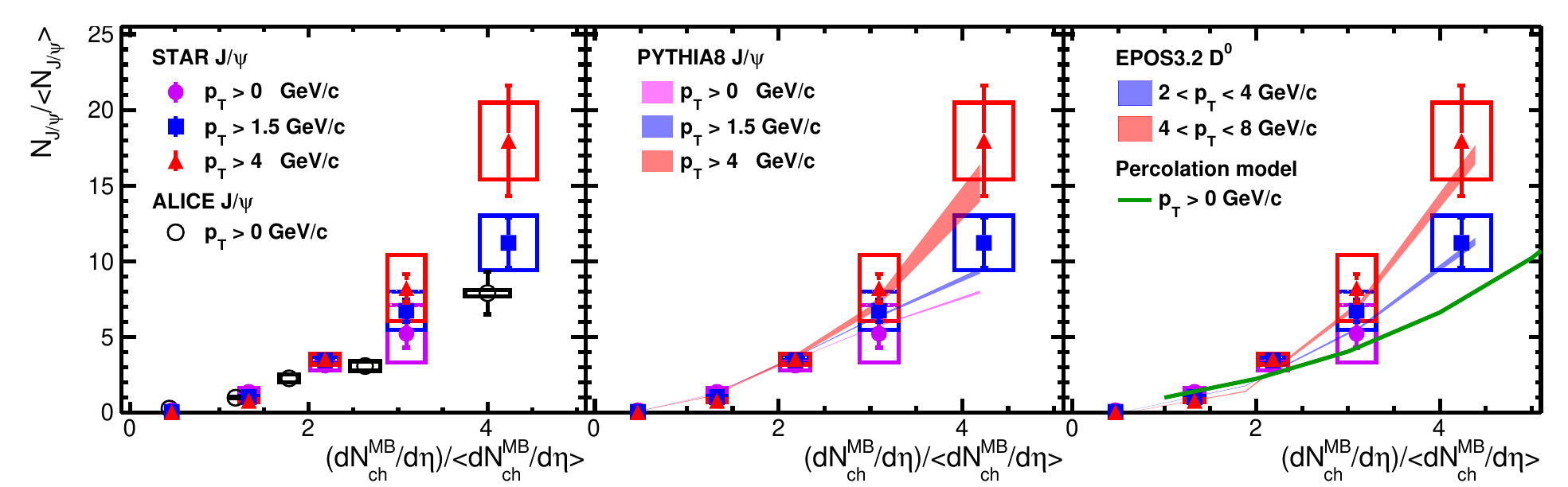}
\caption{(Color online) The multiplicity dependence of \Jpsi production in \pp collisions at $\sqrt{s}=200$ GeV. Purple circles, blue squares, and red triangles represent the results for \Jpsi with \pt greater than 0, 1.5, and 4 GeV/$c$, respectively. Bars and open boxes are statistical and systematic uncertainties, respectively. The ALICE result \cite{ALICEjpsievt} is shown in the left panel. The purple, blue and red bands in the middle panel are generated from PYTHIA8 for \Jpsi with \pt greater than 0, 1.5, and 4 GeV/$c$, respectively. The blue and red bands in the right panel are from EPOS3 model calculations for $D^0$ with $2<p_T<4$ and $4<p_T<8$ GeV/$c$, respectively, while the green curve is from the Percolation model for \Jpsi with $p_T>0$ GeV/$c$. }\label{fig:JpsiEvtAct} 
\end{figure*}

Figure~\ref{fig:JpsiEvtAct} shows the dependence of the $J/\psi$ relative yield on the relative charged-particle multiplicity for \Jpsi $p_T>0$, 1.5 and 4 GeV/$c$, respectively. A strong increase in \Jpsi relative yields with \nch is observed, which seems to be stronger at higher $p_T$ as suggested by data. 
The result for \Jpsi relative yield with $p_T>0$ \gevc in \pp collisions at \sqrts= 200 GeV is compared with that at \sqrts=7 TeV \cite{ALICEjpsievt}.
The two results follow a similar trend despite more than one order of magnitude difference in $\sqrt{s}$, suggesting a weak dependence of the underlying mechanism on $\sqrt{s}$.
Also shown in Fig.~\ref{fig:JpsiEvtAct} are calculations from different MC generators and the percolation model. 
PYTHIA8.183~\cite{pythia8} with the color reconnection scenario describes MPI through pQCD, taking into account the dependence on the energy and impact parameter (the distance between the colliding protons in the plane perpendicular to the beam direction) of \pp collisions. It can describe the \Jpsi relative yield and predicts stronger increase of the \Jpsi relative yield with \nch at higher $p_T$. 
EPOS3 \cite{epos3} uses a Gribov-Regge multiple scattering framework to describe initial \pp collisions and thus includes MPI in both hard and soft processes. Furthermore, it incorporates features of hydrodynamical evolution \cite{epos3hydro} in high-multiplicity \pp collisions, which is suggested to be important for \pp collisions at \sqrts= 7 TeV \cite{ALICED0evt} but has little effect at \sqrts= 200 GeV. Because EPOS3 does not implement $J/\psi$ production, we compare our data with the prediction from EPOS3 on the open charm production. The latter for $2<p_T<4$ ($4<p_T<8$) \gevc is found to be in good agreement with the STAR \Jpsi result for $p_T>1.5$ ($p_T>4$) GeV/$c$. 
The Percolation model~\cite{perco} adapts a framework of color string interactions to describe \pp collisions. In a high-density environment, the coherence among the sources of the color strings leads to a reduction of their effective number. The total charged-particle multiplicity, which originates from soft sources, is more reduced than heavy-particle production for which the sources have a smaller transverse size. The Percolation model prediction is consistent with the data for $p_{T}>0$ GeV/$c$.

\begin{table} \caption{Systematic uncertainties for the measurement of the dependence of $J/\psi$ relative yields on $n_{ch}$. See text for details.\label{tab:JpsiEvtActSys}}
\centering
\begin{tabular}{ccc}

\hline
Type & Uncertainty (\%) \\
         & $\frac{dN_{ch}^{MB}/d\eta}{\langle dN_{ch}^{MB}/d\eta\rangle}$ & $\frac{N_{J/\psi}}{\langle N_{J/\psi\rangle}}$ \\
\hline
Vertex finding  & 2 & 3 \\
Tracking & 4-5 & 4-23 \\
Unfolding & 2-7 & 1-30 \\
Pile-up & 2 & 1-10 \\
\hline
\end{tabular}
\end{table}

\section{Summary}
\label{Summary}

In summary, inclusive \Jpsi production at mid-rapidity $|y| < 1$ in \pp collisions at \sqrts= 200 GeV is studied through the $J/\psi\rightarrow e^+e^-$ channel with the STAR experiment. The measured differential production cross section as a function of \Jpsi \pT can be described within experimental and theoretical uncertainties by CEM calculations for $0<p_T<14$ GeV/$c$, NLO NRQCD calculations for $4<p_T<14$ GeV/$c$, and CGC+NRQCD calculations for $0<p_T<5$ GeV/$c$. 
The total \Jpsi production cross section per rapidity unit times $J/\psi\rightarrow e^+e^-$ branching ratio is $43.2\pm3.0(stat.)\pm7.5 (syst.)$ nb. 
The \Jpsi relative yield is found to increase with charged-particle multiplicity at mid-rapidity. The increase is stronger than a linear rise, and seems to depend on \Jpsi \pT but weakly on the center-of-mass energy of the \pp collisions when compared to the experimental result at \sqrts= 7 TeV. The increase can be described by PYTHIA8 and EPOS3 MC generators taking into account Multiple-Partonic Interactions, and by the Percolation model. 


\section{Acknowledgement}
\label{Acknowledgement}
We thank the RHIC Operations Group and RCF at BNL, the NERSC Center at LBNL, and the Open Science Grid consortium for providing resources and support. This work was supported in part by the Office of Nuclear Physics within the U.S. DOE Office of Science, the U.S. National Science Foundation, the Ministry of Education and Science of the Russian Federation, National Natural Science Foundation of China, Chinese Academy of Science, the Ministry of Science and Technology of China and the Chinese Ministry of Education, the National Research Foundation of Korea, Czech Science Foundation and Ministry of Education, Youth and Sports of the Czech Republic, Department of Atomic Energy and Department of Science and Technology of the Government of India, the National Science Centre of Poland, the Ministry of Science, Education and Sports of the Republic of Croatia, RosAtom of Russia and German Bundesministerium fur Bildung, Wissenschaft, Forschung and Technologie (BMBF) and the Helmholtz Association.






\begin{thebibliography}{9}

\bibitem{Factorization} G.C. Nayak, J.-W. Qiu, G. Sterman, Phys. Lett. B 613 (2005) 45; Phys. Rev. D 72 (2005) 114012; Phys. Rev. D 74 (2006) 074007.
\bibitem{review} A. Andronic et al., Eur. Phys. J. C 76 (2016) 107.
\bibitem{CEM} A.D. Frawley, T. Ullrich, R. Vogt, Phys. Rept. 462 (2008) 125, and R. Vogt private communication (2009).
\bibitem{COCS1} Y.-Q. Ma, K. Wang, K.-T. Chao, Phys. Rev. D 84 (2011) 114001. 
\bibitem{COCS2} M. Butenschoen, B.A. Kniehl, Phys. Rev. Lett.~108 (2012) 172002.
\bibitem{COCS3} B. Gong, L.-P. Wan, J.-X. Wang, H.-F. Zhang, Phys. Rev. Lett.~110  (2013) 042002. 
\bibitem{ALICEjpsievt} B. Abelev et al. (ALICE collaobration), Phys. Lett. B 712 (2012) 165. 
\bibitem{ALICED0evt} J. Adam et al. (ALICE collaobration), JHEP 09  (2015) 148. 
\bibitem{MPI1} P. Bartalini and L. Fano, 1st International Workshop on Multiple Partonic Interactions at the LHC (MPI@LHC08), Perugia, Italy, 27-€"31 October 2008, P. Bartalini and L. Fano eds., DESY, Hamburg (2009), arXiv:1003.4220.
\bibitem{MPI2} T. Sj\"{o}strand and M. van Zijl, Phys. Rev. D 36 (1987) 2019.
\bibitem{MPI3} S. Porteboeuf and R. Granier de Cassagnac, Nucl. Phys. Proc. Suppl. 214 (2011) 181. 
\bibitem{pythia8} T. Sj\"{o}strand, S. Mrenna and P. Skands, Comput. Phys. Commun. 178 (2008) 852.
\bibitem{epos3} H.J. Drescher, M. Hladik, S. Ostapchenko, T. Pierog and K. Werner, Phys. Rept. 350 (2001) 93.
\bibitem{epos3hydro} K. Werner, B. Guiot, I. Karpenko and T. Pierog, Phys. Rev. C 89 (2014) 064903.
\bibitem{perco} E. G. Ferreiro and C. Pajares, Phys.Rev. C 86 (2012) 034903. 
\bibitem{STAR} K. Ackermann et al., Nucl. Instrum. Meth. A 499 (2003) 624. 
\bibitem{STARjpsi} L. Admaczyk et al. (STAR collaboration), Phys. Lett. B 722 (2013) 55; Phys. Rev. C 93 (2016) 064904.
\bibitem{PHXjpsi} A. Adare et al. (PHENIX collaboration), Phys. Rev. D 82 (2010) 012001. 
\bibitem{vpd} W. Llope et al., Nucl. Instrum. Meth. A 522 (2004) 252. 
\bibitem{bbc} J. Kiryluk, AIP Conf. Proc. 675 (2003) 424.
\bibitem{bemc} M. Beddo et al., Nucl. Instrum. Meth. A 499 (2003) 725. 
\bibitem{tpc} M. Anderson et al., Nucl. Instrum. Meth. A 499 (2003) 659. 
\bibitem{tof} B. Bonner et al., Nucl. Instrum. Meth. A 508 (2003) 181. 
\bibitem{embed} B.I. Abelev et al. (STAR collaboration), Phys. Rev. C 79 (2009) 034909. 
\bibitem{Crystal} T. Skwarnicki, DESY F31-86-02 (1986).
\bibitem{CGC2} Y.-Q. Ma, R. Venugopalan, Phys. Rev. Lett.~113 (2014) 192301. 
\bibitem{bfd1} M. Bedjidian et al., arXiv:hep-ph/0311048, 2004.
\bibitem{bfd2} M. Cacciari, P. Nason, R. Vogt, Phys. Rev. Lett.~95 (2005) 122001. 
\bibitem{unfold} G. D'Agostini, Nucl. Instrum. Meth. A 362 (1995) 487.
\bibitem{hftune} L. Adamczyk et al. (STAR collaboration), Phys. Rev. D 86 (2012) 072013. 
\bibitem{zdc} C. Adler et al. , Nucl. Instrum. Meth. A 470 (2001) 488.




\end{thebibliography}



\end{document}